\documentclass[mathleft,fleqn,%
]{an}
\usepackage{graphicx}
\usepackage{natbib}
\usepackage{lscape}
\usepackage[varg]{txfonts}
\def\gas{\mathrel{\hbox{\rlap{\hbox{\lower3pt\hbox{$\sim$}}}\hbox{\raise2pt\hbox{$>$}}}}}
\def\las{\mathrel{\hbox{\rlap{\hbox{\lower3pt\hbox{$\sim$}}}\hbox{\raise2pt\hbox{$<$}}}}}
\begin{document}

% The following seven commands are intended for editorial usage and
% should be ignored by the author(s).
\Pagespan{1}{}% Document's page range. 
% If second parameter is left empty, the last page is computed
% automatically.
%\Yearpublication{2016}%
%\Yearsubmission{2016}%
%\Month{7}%   
%\Volume{999}%  
%\Issue{0}% 
%\DOI{asna.201400000}% 

\title{V-type Near-Earth asteroids: dynamics, close encounters and impacts with terrestrial planets\thanks{Data
from Dr. Mattia A. Galiazzo}}

\author{M.\,A.  Galiazzo\inst{1,2,3}\fnmsep\thanks{M. A. Galiazzo:
        {mgaliazz@uwo.com / mattia.galiazzo@gmail.com} }
% Example for footnote, note the usage of the \texttt{fnmsep} command
% as separator between institute number and footnote mark}
E.\,A. Silber\inst{1,2}\and  D. Bancelin\inst{3,4}
}
\titlerunning{V-NEAs: dynamics, close encounters and impacts with terrestrial planets}
\authorrunning{M.\,A. Galiazzo, E.\,A. Silber \& D. Bancelin}
\institute{
Department of Physics and Astronomy, The University of Western Ontario, London, Ontario, N6A 3K7, Canada
\and 
Centre for Planetary Science and Exploration (CPSX), London, Ontario, Canada, N6A 3K7
\and
Institute of Astrophysics, University of Vienna, Turkenschanzstr. 17, A-1180 Vienna, Austria
\and 
IMCCE, Paris Observatory, UPMC, CNRS, UMR8028, 77, Av. Denfert-Rochereau F-75014 Paris, France}

\received{Jul. 1, 2016}
\accepted{XXXX}
\publonline{XXXX}


\abstract{%
 Asteroids colliding with planets vary in composition and taxonomical type. Among Near-Earth Asteroids (NEAs) are the V-types, basaltic asteroids that are classified via spectroscopic observations. In this work, we study the probability of V-type NEAs colliding with Earth, Mars and Venus, as well as the Moon.
 We perform a correlational analysis of possible craters produced by V-type NEAs. To achieve this, we performed numerical simulations and
 statistical analysis of close encounters and impacts between V-type NEAs and the terrestrial planets over the next 10 Myr.
 We find that V-type NEAs can indeed have impacts with all the planets, the Earth in particular, at an average rate of once per $\sim12$ Myr.
 There are four candidate craters on Earth that were likely caused by V-type NEAs.  
 % look at the World Wide Web (URL
  %{http:/\slash{}www.aip.de\slash{}AN/}) where this text and
 % accompanying class files can be obtained from.}
}
\maketitle

\section{Introduction}\label{sec: intro}

 V-type Near-Earth asteroids (V-NEAs) are basaltic asteroids with a perihelion less than 1.3 au.
 Dynamically speaking, these objects exist as NEAs in a chaotic region; however, they are peculiar, because they originated from specific asteroid families, mainly from the Vesta family \citep{Sea1997,Car2007,Del2012,Gal2012} in the main asteroid belt, even though some V-type asteroids can be found in other families (Eunomia, Magnya, Merxia, Agnia, Eos and Dembowska) \citep{Car2014,Hua2014}, albeit in significantly smaller numbers. 
Compared to other NEAs, V-NEAs can have different orbits, size distribution and, of course, composition (being only basaltic asteroids coming from peculiar bodies, e.g. 4 Vesta). In case of impacts, V-NEAs can produce a wide range of crater sizes and impact material (e.g., this was shown in part, with some V-NEAs in \citet{Gal2013b}).
 Some of V-NEAs are in a resonance, e.g. (7899) 1994 LX and (137052) Tjelvar in M4:3 (mean motion with Mars),  (1981) Midas (which is also the largest V-NEA) in J3:2 (mean motion with Jupiter),
and (4688) 1980 WF in the $\nu_6$ secular resonance, which creates one of the most unstable interactions (see \citet{Mig1998}, \citet{Bot2002}).
Close planetary encounters strongly influence the dynamics of bodies in NEA space.
 The dynamics in the NEA region is therefore the result of a complicated interplay between resonant dynamics and close encounters \citep{Mic2005}. 
 Known V-NEAs
are currently confined in these ranges of osculating elements (present semi-major axis $a_0$, eccentricity, $e_0$ and inclination, $i_0$, as listed in Table \ref{single}):
 $0.825< a_0 <2.824$ au (between resonances with the Earth, E4:3, and with Jupiter, J5:2),
 $0.29<e_0<0.90$ and $2^\circ<i_0<49^\circ$.
Furthermore, V-NEAs typically have a semi-major axis $a\las1.8$ au and $e\gas0.4$. This is different from, for example, the Hungarias E-type NEAs 
 which preferentially have $1.92 < a < 2.04$ au and $0.32 < e <0.38$, as shown in \citet{Gal2013}
 
\begin{figure}
\centering
 \centering \includegraphics[width=0.42\textwidth]{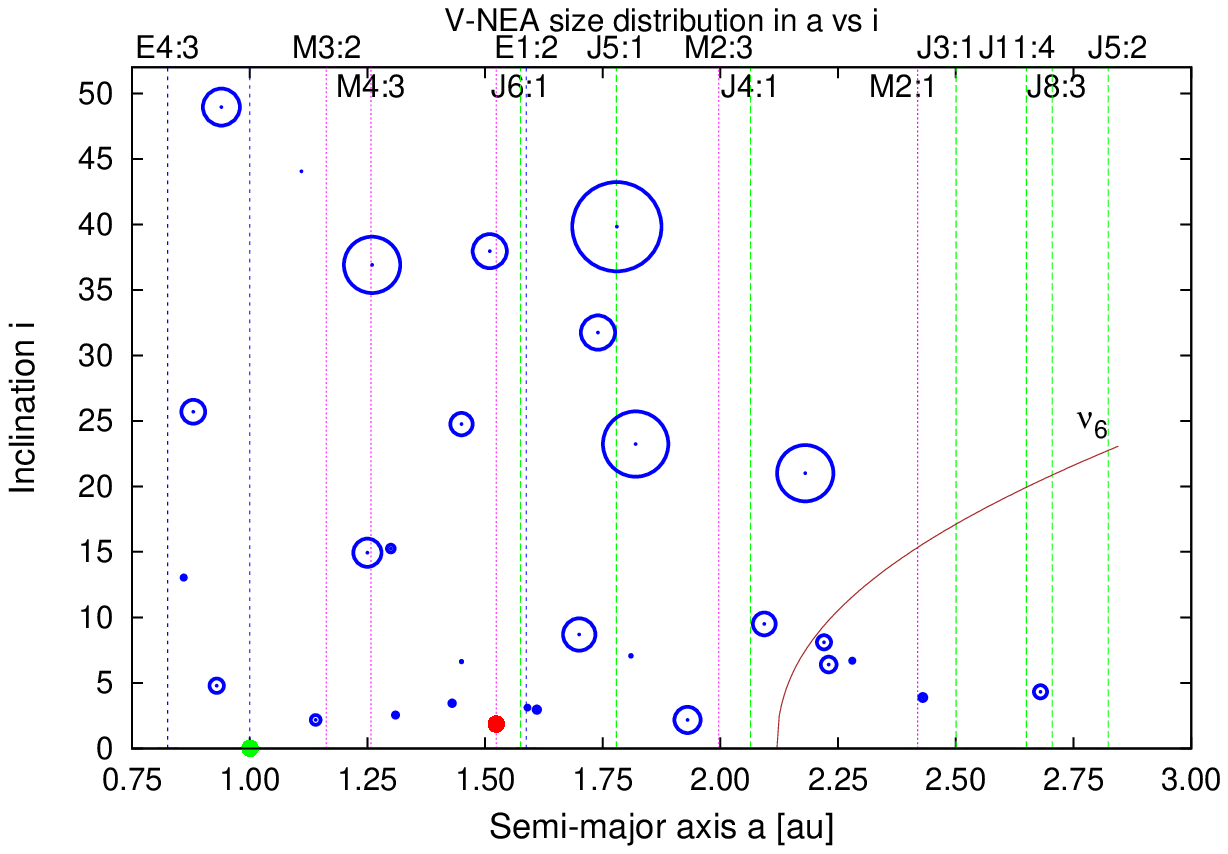}
 \centering \includegraphics[width=0.42\textwidth]{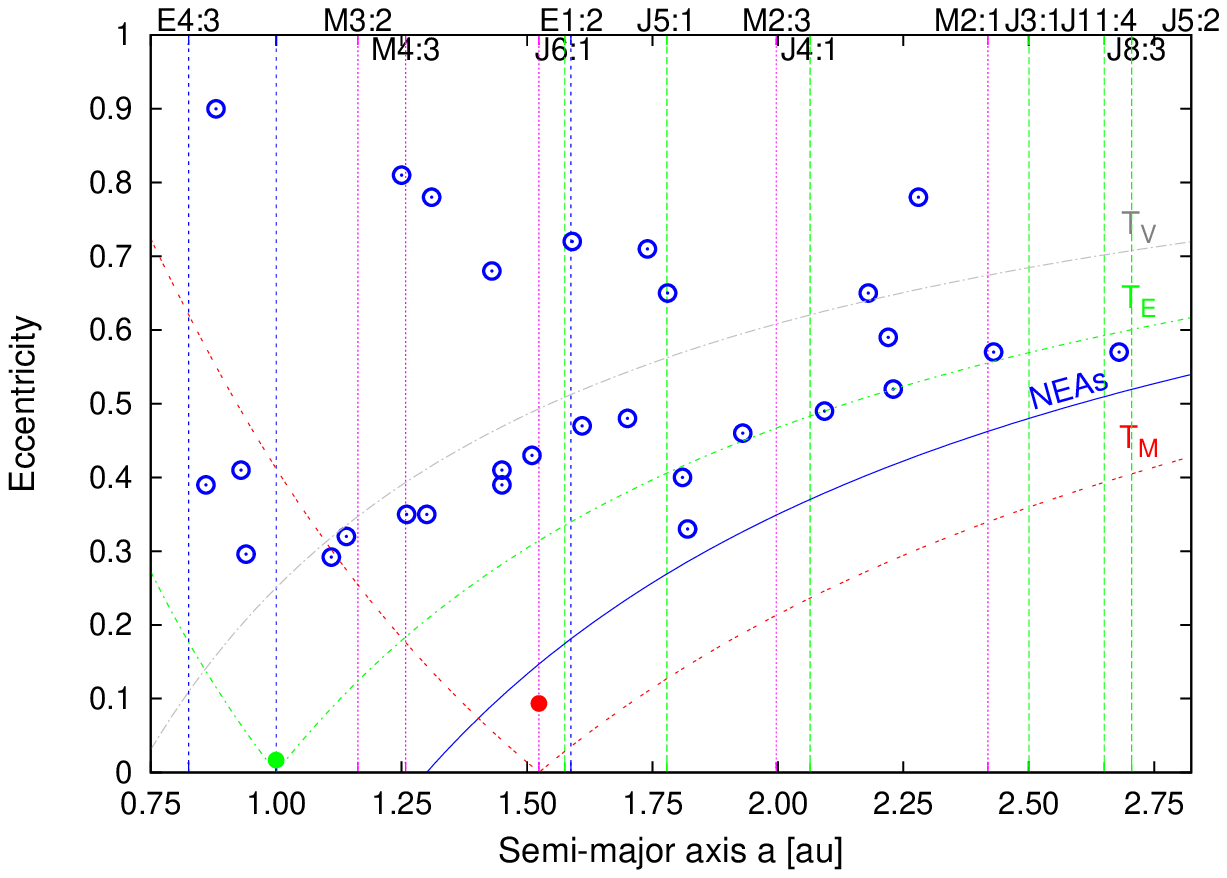}
\caption{Location and size distribution of the V-NEAs in the ($a$,$i$) and ($a$,$e$) plane.
 Circles (in the upper plot) are proportional to the V-NEA size: the smallest asteroid is 62 m in diameter
 and the largest is 3.4 km. The blue, violet and green vertical lines represent the mean motion resonances with Earth, Mars and Jupiter,
 respectively. The green dot is Earth and the red dot is Mars. The semi-dotted curves are the Tisserand curves for the respective terrestrial planets.}
	\label{Vlocation}
\end{figure}

 Figure ~\ref{Vlocation} 
 shows the ($a$,$i$)
 (with the size distribution) and ($a,e$) plane for V-NEAs, 
 where it is evident that the largest asteroids reside at high inclinations, $i\gas20^\circ$.
Their size ranges from $\sim$60 m to $\sim$3.1 km. These quantities are computed using a standard\footnote{The average of
 the known NEAs: 0.33.} albedo for basaltic asteroids and their absolute magnitude \citep{Ted1992}.

This work aims to investigate if present day basaltic or V-type Near-Earth Asteroids (V-NEAs) can impact terrestrial planets
 (specifically, Venus, Earth and its satellite, and Mars) within the time span of 10 million years (Myr). 

% ----------------------------------------------------------
  Furthermore, we aim to perform the following: 
\begin{description}
\item[(i)] establish if the probability of close encounters, the impact probability for each V-NEA
 (assuming the number of impacts among all its clones), the rate of impacts, and the size of craters
 produced by V-NEAs that collide with planetary surfaces. 
\item[(ii)] examine if some asteroids could impact planetary surfaces at high velocities. 
Compared to slow velocity objects, high velocity asteroids can generate tramendeous amount of kinetic energy upon
 the collision with a planetary surface. Such impacts might cause catastrophic events, similar to disruptive events
 usually made by comets \citep{Jef2001}. 
\item[(iii)] perform a statistical analysis of close encounters and impacts between present (observed and synthetic populations of) V-NEAs and the terrestrial planets. For impacts, we primarily aim to establish the impact energy and size of impact craters on planetary surfaces,
 particularly the Earth. We also aim to identify possible candidate craters on the Earth, using past geological studies on the material of the impactor, in addition to our constratins on the crater sizes obtained through our results coming from orbital simulations and crater formation simulations.
\end{description}
 % ----------------------- end bulleted section ---------------------------------
  
 We performed numerical computations of orbits and also simulated crater formation on planetary surfaces.\\
The paper is organized as follows: Section~\ref{model} describes the model, subdivided in two subsections, one for the numerical simulation of the impacts and one for the simulation of the orbital evolution of the V-NEAs. In Section~\ref{results} we revisit the numerical results, and
 finally, our conclusions are given in Section~\ref{finallyMario}.

\section{Model}

\label{model}

\subsection{Numerical simulations of the V-NEAs orbits and analysis}
The orbital evolution of each V-NEA, including its 49 clones, has been computed using its real orbit. The clones of each asteroid were generated with these elements: 
 $a = a_{0}\pm  0.005$, $e = e_{0} \pm 0.003$ and $i = i_{0}\pm 0.01$,
 where elements with $0$ are the initial osculating elements
 of the V-NEAs. This range was selected based on \citet{Hor2003a}, who
showed that the aforementioned intervals of osculating elements are 
reasonable to perform orbital simulations for minor bodies in such
 chaotic orbits. 
 Therefore we propagated 1450 V-NEAs orbits.
 Each clone  was integrated over the time span
 of 10 Myr into the future using the Lie integrator \citep{Han1984,Egg2010,Ban2012} with an accuracy parameter set to $10^{−13}$. The code was upgraded by a subroutine designed to compute encounter velocity, deflection of the orbit, impact angle, impact velocity
 after an impact is identified (see \citet{Gal2014}).
 Orbits were taken from the HORIZONS Web-Interface \footnote{http://ssd.jpl.nasa.gov/horizons.cgi\#top} at Epoch 2453724.5 JDTBT
 (Epoch Julian Date, Barycentric Dynamical Time).
NEAs with a V-type spectrum were taken from \citep{Xu1995,Bin2001,Bus2002,Dan2003,Bin2004,DeM2009}. The diameters of the
 bodies in case of impacts were obtained from The Near-Earth Asteroids Data Base \footnote{http://earn.dlr.de/nea/} (NEADB). If the diameter was not available, we used the equation from \citet{Ted1992}, assuming that the average albedo $\rho_V=0.33$ for the
 known V-NEAs (NEADB):
$D=\frac{1329}{<\rho_V>}10^{-\frac{H}{5}}$. $H$ is the absolute magnitude, $D$ the diameter,
and $\rho_V$ is the average albedo.
 Only gravitational perturbations caused by the planets from Mercury to Neptune were taken into account. We neglected the contribution of non-gravitational effects mainly Yarkovsky and YORP\footnote{Yarkovsky–O'Keefe–Radzievskii–Paddack effect, which is a secondary effect on Yarkovsky because it influences the spin-axis orientation} effects. Indeed, within our
 integrational timescale, the semi-major axis drift caused by the Yarkovsky
 effect to the NEAs will be very weak as $da/dt=10^{-3}-10^{-4}$ au Myr$^{-1}$ (Farnocchia et al., 2013), and it is negligible compared to the 
 effects of the close encounters with planets. In addition, close encounters are very frequent for the V-NEAs. Thus, our statistical results will not be affected within 10 
 million years of integration.

 The code searches for impacts when asteroids have close encounters with a planet. A close encounter occurs when an asteroid has a
 planetocentric distance less than 0.0025 au (= 1 LD, one lunar distance) from the Earth, and for other bodies scaled to the relative Hill sphere
 \citep{Gal2013}. 
An impact at the surface of the planet should occur if (a) an asteroid crosses the radius distance of the planet and (b) when the diameter of the body is larger than 
the critical diameter computed via eq. (7) of \citet{Col2005}, which accounts for atmospheric pressure (e.g. relevant on Venus).
 In case condition (a) is valid, but not (b) we consider the body to burn up as a meteor.
 For each detected planetary impact, we derived the impact velocity, and impact angle
 \citep{Gal2014,Gal2013}.
 Where necessary, we took into consideration the atmospheric effect as modeled by \citet{Col2005}.
 For example, the impact velocity for Venus is significantly reduced at surface as a result of the atmospheric pressure.

 Additionally, we used the derived impact velocities and impact angles for the purpose of hydrocode modeling of impact craters.
 Finally, we derived statistical distributions of the impact velocities and crater sizes, which are compared with the real crater sizes on
 the terrestrial planets, giving a special consideration to a cutoff size for the V-NEA-produced craters.

\subsection{Hydrocode modeling
of impacts}

For numerical simulations of impact crater processes we employ iSALE-2D, a multi-material, multi-rheology shock physics hydrocode
 \citep{Col2004,Wun2006}. We use the ANEOS equation of state (EoS) for basalt to approximate the V-type asteroid material,
 as well as the crustal material of Venus. For the Earth, we use granite \citep{Pie1997} to represent the crust.
 All simulations include the material strength \citep{Col2004} and damage \citep{Iva2010} models, as well as the effects of
 acoustic fluidization \citep{Mel1979}. Since vapour production is not relevant in this study, all material with a density $>$300 kg/$m^3$
 was removed. Impact velocities, derived from evolution and orbital dynamics of V-type particles, are scaled to the vertical component only,
 due to the cylindrical symmetry of iSALE-2D. The high resolution zone cell size is varied according to the size of the impactor,
 keeping the number of cells per projectile radius constant at 10, as this value offers a reasonable trade-off between computational time
 and resolution errors \citep{Pot2013}.

\noindent

%%%%%%%%%%%%%%%%%%%%%%%%%%%%%%%%%%%%%%%%%%%%%%%%%%%%%%%%%%%%%%%%%%%%%%%%%
%%%%%%%%%%%%%%%%%%%%%%%%%%%%%%%%%%%%%%%%%%%%%%%%%%%%%%%%%%%%%%%%%%%%%%%%

%%%%%%%%%%%%%%%%%%%%%%%%%%%%%%%%%%%%%%%%%%%%%%%%%%%%%%%%%%%%%%%%%%%%%
\section{Results}
\label{results}

\subsection{Dynamics and close encounters}

 We find that on average, a V-NEA can pass 54 times at less than 0.0025 au in planetocentric distance with
 the Earth in 10 Myr, well within the orbital space of Potential Hazardous Asteroids
 (PHAs)\footnote{PHAs are NEAs whose Minimum Orbit Intersection Distance (MOID)
  with the Earth is 0.05 au or less and whose absolute magnitude is
 $H\leq 22.0$}. 
Based on our studies, 90.6\% of the V-NEAs are possible PHAs.
 Furthermore, if we scale (see \citet{Gal2013})
 this distance to the Hill Sphere of the other planets,
then we obtain  75.2\% for Venus, 79.3\% for Mars and 57.9\% for the Moon.
This means that nearly all the clones encounter the Earth, $\sim$9 out of 10 (see Table ~\ref{impactcom}) and on average,
 the number of close encounters with the Earth is higher than with the other planets. The Moon has fewer close encounters than the 
 Earth (in accordance with other results for all the NEAs, see e.g. \citet{Jef2001,Iva2008},
 and it is the least encountered overall.
Considering their relative speed, typical encounter velocities for the terrestrial planets vary and are shown in Table\ref{cratercom}.
 The encounter velocities for
 Venus are typically higher, as expected from other work (e.g.  22.2 km/s, \citet{Iva2008}).

\subsubsection{Single cases}
Considering the asteroids singularly, all V-NEAs have a certain probability to encounter each planet, except Magellan. Magellan is the only asteroid which has a very low probability to encounter any planet, with no clones ever coming to the proximity of Venus. In fact, its initial MOID with Venus is the highest among the V-NEAs (see Table \ref{single}). This is
 due to its high inclination (far from resonances)
 and high perihelion (the highest among the sample).
In fact, it does not initially cross the Earth's orbit, and therefore there is
 no increase of its eccentricity due to planetary close encounters.

\subsection{Impacts}
In addition to close encounters, V-NEAs can have impacts with all
 the planets considered in our study (Venus, Earth and Mars) and the Moon
 as shown in Table \ref{single}, \ref{impactcom} and \ref{cratercom}.
We found a relationship between the number of impacts and the MOID: impacts are possible only if the MOID with the mutual planet
 is maximum 0.3 au and there is an exponential distribution between the number of impacts and the MOID up to MOID 0.05 au.
 We note that Mars and Moon have the lower probabilities of 
experiencing an impact.
In particular, the asteroids which have the highest probability to collide with a planet in the next 10 Million years are:
 1992 FE with Venus, 1996 EN and 2000 HA$_{24}$ with the Earth, and 2003 FU$_3$ with the Moon (Table \ref{single}).
The average impact velocity with the Earth, as considered before the atmospheric effect becomes relevant
 (i.e. entry velocity at 100 km altitude), is somewhat lower than that found by \citet{Gal2013b},
 and higher than the velocity determined by \citet{Iva2008}. However, it is well within the standard deviation.
The average impact velocities are as follows:  
 $v=20.0\pm7.3$ km/s, $v=19.3$ km/s and $v=\sim 21$ km/s, where the values represent the average velocities found in this work
 (the V-type NEA population), \citet{Iva2008} (NEAs with $H<17$) and \citet{Gal2013} (all the NEAs), respectively.
Our simulations suggest that average impact velocities are also
  higher for Mars (e.g. see \citet{Iva2008}). However, we note that this particular
 result could be a statistical bias because we found only 3 impacts with Mars. 
Nevertheless, if this assertion is correct, then it follows that on average,
 V-NEAs will produce larger craters than that created by NEA populations with
 slower velocity distribution (due to difference in their kinetic energies).

\begin{table*}
%\begin{center}
\caption{Impact  probability (in percentage) per 10 Myr and MOID of the present osculating elements
 for each planet and each V-NEA, including its size ($D$) and initial osculating elements
 ( $a_0$ $e_o$ $i_0$). $S_i$ and $E_s$, respectively, and the probability of impact with the Sun and the probability of escaping.
 $V_i$,$E_i$,$Mo_i$ and $Ma_i$ stand for impact probability (in percentage) for Venus (V), Earth (E), Moon (Mo) and Mars (Ma), respectively. $MO$ stands for MOID.
 The small numbers in parenthesis mean possible meteors.}
\begin{tabular}[h]{lccccccccccccc}
\hline 
\hline
V-NEAs  & D[km] &$a_0$& $e_o$& $i_0$     & V-MO  & E-MO & Ma-MO & V$_i$ & E$_i$\  & Mo$_i$ &Ma$_i$ & S$_i$ & Es\\ %& \% (E/V)& \%(E/T)& \% (V/T) \\
\hline
Midas   & 2.20 & 1.776& 0.65& 4.0 &  0.1807 & 0.0038 & 0.0269  & 	 0         &   0          &       0        &    2        &         0      &  0 \\
1996 AJ$_1$ &0.20 &  1.310& 0.78& 2.5    &    0.0014 &     0.0048   &0.0139 &    4$_{(4)}$         &   0          &       0      &    0        &    52          &  0 \\       
1997 GL$_3$& 0.08&2.276 & 0.78 & 6.7& 0.0556 & 0.0014 &  0.0276  &   0   &    0  &0 & 0&66 & 32  \\
1999 RB$_{32}$&0.30 &2.434& 0.57& 3.9& 0.3314 & 0.0621 &   0.0127 & 0 & 0 & 0 &  0  & 52 &  42 \\
2000 DO$_1$&0.20 &1.430& 0.68& 3.5 & 0.0614 & 0.0135 & 0.0202        & 0$_{(2)}$ & 4 & 0 &  0  & 56 & 36 \\
2001 PJ$_{29}$&0.06 &1.449& 0.39& 6.6& 0.1952 & 0.0235 & 0.0659      & 0$_{(2)}$ & 6 & 0 &  0  & 10 & 6  \\  
2003 FT$_{3}$  &0.47 &2.681& 0.57 & 4.3 & 0.4397 &  0.1646 &      0.0129   & 0 & 0 & 0 &  0  & 26 & 44 \\
2003 FU$_{3}$  &0.15 &0.858& 0.39& 13.0 & 0.1395 &  0.0819 &      0.4502   & 0$_{(2)}$ & 0 & 4 &  0  & 12 & 4  \\
2003 GJ$_{21}$&0.06 &1.811& 0.40 & 7.1 & 0.3673 &  0.0890 &      0.1215   & 0 & 4 & 2 &  0  & 32 & 6  \\
2004 FG$_{11}$ &0.15 &1.588& 0.72 & 3.1&  0.0180 &  0.0210 &      0.0032   & 0 & 2 & 0 &  0  & 34 & 18  \\
2008 BT$_{18}$ &0.51 &2.222& 0.59 & 8.1&  0.1873 &  0.0110 &      0.1569   & 0 & 0 & 0 &  0  & 68 & 28  \\ 
Verenia     &0.87 &2.093& 0.49 & 9.5 & 0.3470 &  0.0719 &      0.2323   & 0 & 2 & 0 &  0  & 60 & 14  \\   
Nyx     &0.50 &1.928& 0.46 & 2.2 & 0.3217 &  0.0563 &      0.0139   & 0 & 6 & 0 &  0  & 50 & 10 \\    
Magellan     &2.78 &1.820& 0.33& 23.3 & 0.5233 &  0.2373 &      0.1135   & 0 & 0 & 0 &  0  & 0  & 0 \\
1980 WF     & 0.60&2.234& 0.52 & 6.4 & 0.3693 &  0.1141 &      0.1153   & 0 & 0 & 0 &  0  & 74 & 24 \\
Sekhmet     &1.41 &0.947& 0.30& 49.0 & 0.0192 &  0.1116 &      0.3718   & 0 & 6 & 0 &  0  & 0  & 0 \\
1992 FE     &0.97 &0.929& 0.41&  4.7&  0.0060 &  0.0335 &      0.2958   & 14& 0 & 0 &  0  & 6  & 2 \\
1993 VW     &1.16 &1.696& 0.49&  8.7&  0.2248 &  0.0607 &      0.0104   & 0 & 2 & 0 &  0  & 34 & 6 \\
1994 LX     &3.10 &1.261& 0.35& 36.9&  0.1118 &  0.1545 &      0.2556   & 0 & 4 & 0 &  0  & 0  & 2 \\
1996 EN     &1.57 &1.506& 0.43& 38.0&  0.2504 &  0.0231 &      0.0550   & 0 & 8 & 0 &  0  & 6  & 2 \\
Tjelvar   &1.00 & 1.248& 0.81& 14.9 & 0.0646 &  0.0686 &      0.0007   & 0 & 0 & 0 &  2  & 10 & 0 \\
2000 BD$_{19}$   &0.97 &0.876& 0.90& 25.7&  0.0214 &  0.0904 &      0.2944   & 4 & 6 & 0 &  0  & 0  & 0 \\
2000 HA$_{24}$   &0.35 &1.140& 0.32&  2.2&  0.0571 &  0.0274 &      0.0209   & 6 & 8 & 0 &  0  & 10 & 6 \\
2003 YQ$_{17}$   &2.02 &2.180& 0.66& 21.0&  0.1144 &  0.1055 &      0.2659   & 2 & 0 & 0 &  0  & 56 & 20  \\
1999 CV$_8$   &0.28 &1.297& 0.35& 15.3 & 0.2155 &  0.0571 &      0.1533   & 4 & 6 & 0 &  0  & 8  & 0 \\
2003 EF$_{54}$   &0.23 &1.609& 0.47&  3.0&  0.1221 &  0.0421 &      0.1029   & 6 & 4 & 2 &  2  & 24 & 10 \\
1998 WZ$_6$   &0.80 &1.452& 0.41& 24.8 & 0.2438 &  0.0341 &      0.0743   & 2 & 6 & 0 &  0  & 10 & 2 \\
2003 EG   &1.47 &1.738& 0.71& 31.8 & 0.1911 &  0.3616 &      0.2311   & 0 & 0 & 0 &  0  & 44 & 6 \\
 2003 YT$_1$ &1.10 &1.110& 0.29& 44.1 & 0.2446 &  0.0025 &      0.3813   & 0 & 6 & 0 &  0  & 0  & 0 \\
\hline
\end{tabular}
%}
	\label{single}
%\end{center}
\end{table*}

Concerning the probability of an impact,
 we assume the average impact probability per body per Gyr 
 (Table ~\ref{impactcom}) and we compare the probability of impacts to that of the Hungarias (the Hungaria
 family is one of the principal main belt sources of NEAs \citep{Bot2002,Gal2013,Gal2014})
 and to observed planetary crossers with $H<17$ ($OPC17$).
The average impact probability per 1 Gyr is:
\begin{equation}
N_{imp,Gyr}=\frac{N_{Tot}}{N_b}t_s
\end{equation}
 where, $N_{Tot}$ is the total number of impacts found in the simulations, $N_b$ is the total number of integrated bodies (1450)
 and $t_s$ is the time scaling factor ($t_s=100$, 1 Gyr divided per the integration time 10 Myr).

The V-NEAs undergo a higher number of collisions with the Earth (about $\sim3$\% of the population\footnote{Assuming the observed population
 and the clones as a sample of the entire basaltic NEA population} in 10 Myr).

The contribution of the V-types to Mars as impactors is similar to what has been found for the $OPC17s$,
 but lower than for E-types/Hungarias, that is, considering that Hungarias are
the major source of E-types. Although E-types might come from other sources, Hungarias are still about 60\% E-type or achondritic enstatite \citep{War2009}).
The impact rates with Venus are half that of the Earth for the V-NEAs. We note however, that in addition, we also 
 find that each V-NEA in the entire population has a probability of 0.3\% to become a meteor for this planet in 10 Myr.

 Comparing our work to that of \citet{Cha2004}, we find that V-NEAs impactors can generate up to 5-6 Mt of impact energy at the
 point where the curve of the plot becomes steeper. \citet{Cha2004} showed that this kind of impact happens approximately
 once every 10 to 100 million years. Our results demonstrate, considering the percentage of impacts,
 that V-NEAs can produce such an impact slightly
 more frequently than once every 10 million years, which is  in a reasonable agreement with \citet{Cha2004}.
 In our study, there are 0.84 asteroids
 ($2.897$ \% out of 29, see Table ~\ref{impactcom}) that can collide with the Earth in 10 Myr, thus scaling in time,
 this implies that there is one impact every 11.9 Myr (see Eq.~\ref{impactrate}).
 However, we underline the fact that this is applicable only to V-type asteroids. Furthermore, since there are at least 6 other types
 (i.e., S-types NEAs are more numerous than V-types \citep{Bin2002}), this could mean that impacts of this size range could happen
 more frequently in the last 10 Myr at least.

The time lapse of impacts with respect to a given planet is:
\begin{equation}\label{impactrate}
t_{IRP}=\frac{t_{int}}{p_{IV} n_A}
\end{equation}
where $p_{IV}$ (see Table ~\ref{impactcom}) is the percentage of V-NEAs impacting a relative planet during
 the time ($t_{int}$) in which their orbits are propagated, and $n_A$ is the number of considered asteroids (in our case 29).
 Therefore, $t_{IRP}$ is 21.7 Myr for Venus, 125.0 Myr for the Moon, and 166.7 Myr for Mars.

\begin{table}[!h]
\begin{center}
\caption{Impact rate of V-NEAs compared to all the Terrestrial planet-crossing asteroids and number of close encounters, for Venus, Earth, Moon and Mars.
The values from left to right are: the considered planet, the percentage of V-NEA clones which have close encounters with the planet, average number of close encounters (CEs) per V-NEA, average collision probability per one body V-type NEA, E-types (Hungarias) (from \citet{Gal2013} and asteroids (planetary crosser) with $H<17$ (from \citet{Iva2008}, per Gyr.). 
* in the Hungarias' study, the Earth was considered together with the Moon in a barycentric system.
** this result comes from a synthetic population of NEA that evolved 
from the original Hungarias in the main belt.
}
\begin{tabular}{lccccc}
\hline 
\hline
Planet &\%CEs &$\bar{\#}_{CE}$ &$N_{imp,Gyr,V}$   &  $N_{imp,Gyr,E}$** & $A_{H<17}$\\
\hline
Venus  &75.2 & 40  &1.586	   & 0.014      & 4.500       \\ 
Earth  &90.6  & 54  &2.897	   & 0.004*     & 3.400      \\ 
Moon   &57.9 & 3   &0.276	   &   -        & 0.160      \\   
Mars   &79.3 & 11  &0.207	   & 0.006      & 0.210      \\   
\hline
\end{tabular}
\label{impactcom}		
\end{center}
\end{table}

\subsection{Impact craters}
\begin{figure*}[!h]
  \caption{Map of the confirmed craters on Earth. Red asterisk indicate confirmed
    or suspected basaltic impactors. Remaining craters are shown as blue circles (diameter less than 35 km) or orange squares (greater than 35 km). 
Data from  Goderis, Paquay, \& Claeys (2012) and cited papers in Section 3.}
\includegraphics[width=6.5in,angle=0]{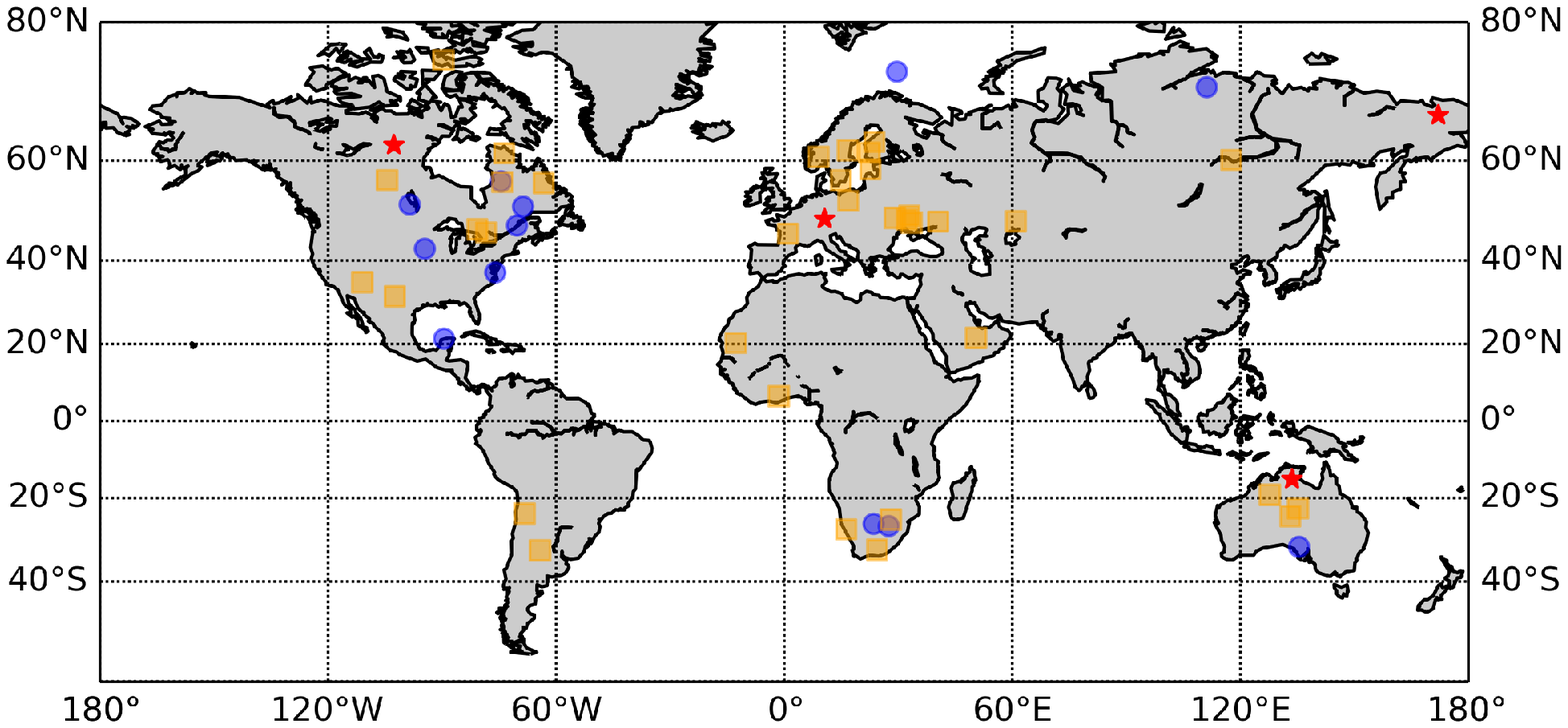}
	\label{map}
\end{figure*}

Because our sample of the asteroids are in meter to kilometer size range, 
 we now deal with impact craters, assuming a simple case of single body impacts
 (e.g. no break up during their passage throught the atmosphere of Venus
 and the Earth).
Table 3 shows the impact  crater results (including impact rate time, crater diameter and energy released during the impact).
For the Earth we attempted to determine the most probable crater candidates produced by the V-NEAs, taking into account their sizes
 and the material of the impactors, especially where it was available.
The size of the V-NEAs' craters expected from our simulations range from 0.7 km to 32.7 km,
 thus we concentrated on the Earth's craters with a $D< 33$ km.
Our results suggest that the craters most likely produced by V-NEAs on
 the Earth (shown with an asterisk in Figure~\ref{map})
  are known or thought to have been generated by a basaltic impactor
 (under past geological studies),
 with a diameter within the range found by impacts with V-NEAs in our simulation.
The two known basaltic impact structures are the $<400$ Myr old Nicholson crater in Canada
 (62$^\circ$N, 102$^\circ$W; 12.5 km diameter) and
 the 646 Myr old Strangways crater in Australia (15S, 133WE; 25 km diameter) 
 \citep{God2012}.
Two other craters whose impactor's composition has not yet been confirmed (although the preliminary analyses suggest that the basaltic composition is possible) could also have originated from V-NEA impactors: 
  the 3.5 Myr old crater El'gygytgyn in Russia (67$^\circ$N, 172$^\circ$E),
  and the 15 Myr old Ries crater\footnote{Concerning the impactors, for the Ries crater, an aubrite-like (achondritic)
asteroid has been proposed \citep{Mor1979,Hor1983,Per1987,Tag2006,Art2013}. For El’gygytgyn, the distribution of impactor traces in the continuous
 impactite section \citep{Val1982,Wit2012} and an apparent Cr enrichment \citep{Gur2004} may validate
previous reports of an achondritic (ureilite?) projectile.} in Germany
 (48$^\circ$N, 10$^\circ$E; 24 km diameter).
The craters marked with blue circles have diameters less than 35 km, and those larger than 35 km are marked with orange squares.

\begin{table*}[!h]
\begin{center}
\caption{V-NEAs crater sizes among the planets (D), impact velocity ($v_{imp}$), close encounter velocity ($v_{CE}$), at 1 $LD$, the impact rate time ($\bar{t}_{IRP}$) and impact energy (En.). All values
with their average (first line), and
 maximum and minimum diameter (second line). \#I is the number of the impactors (per relative planet) on which the statistics was done.}

\begin{tabular}{lccccc}
\hline 
\hline
Planet& $\bar{t}_{IRP}$ [Myr]  & D[km] &$v_{CE}$ &$v_{imp}$ & En.[Mt]  \\ 
\hline
 \#I      &   & Ave.  & Ave.  & Ave.  & Ave.   \\
    &   &  $Min\quad  Max$ & $Min \quad Max$   & $Min\quad  Max$ & $Min\quad  Max$ \\
\hline
Venus  &  21.7   &  $7.9\pm5.3$        & $23.1\pm8.5$  &$10.8\pm7.4$ &   $(0.4\pm 0.7)\times10^5$            \\
23 & &   $1.3\quad  16.6$	   & $2.9 \quad 74.2$  &$1.9 \quad 25.4$  & $8.16 \quad 3.15\times10^6$       \\
\hline
Earth & 11.9 &  $11.1\pm7.9$ 	   &$19.1\pm6.9$ &$20.0\pm7.3$&$(2.3\pm 0.6)\times10^5$       \\
42 & &   $0.7\quad  32.7$	   &$2.0 \quad 60.8$&$8.9 \quad 34.0$&  $2.79 \quad 2.78\times10^6$    \\
\hline
Moon & 124.9 &   $3.4\pm1.2$	   &$18.9\pm6.6$ &$15.4\pm1.1$& $(1.9\pm 2.0)\times10^2$                \\
4 & &   $2.0\quad  6.3$	   &$3.1 \quad 44.9$ &$14.8 \quad 17.0$& $ 13.0 \quad 4.84\times10^2$    \\
\hline
Mars &166.6  &  $50.3\pm27.8$  	   &$15.4\pm4.1$&$22.0\pm6.0$& $(4.3\pm 6.5)\times10^5$           \\
3 & &   $20.3\quad  75.1$	   & $0.6 \quad 43.0$&$15.1 \quad 26.0$& $ 501 \quad 1.18\times10^6$    \\
\hline
\end{tabular}
	\label{cratercom}	
\end{center}
\end{table*}

The kinetic energy released during the largest impact examined here can reach up to
 3 Mt of TNT equivalent  (1 Mt of TNT $=  4.184\times10^{15}$ J).
 This amount of energy is $\sim2\times10^8$ greater than that released over Hiroshima, and $\sim6\times10^6$ greater than produced by the
 Chelyabinsk event \citep{Bro2013}.

 An impact of such scale could cause regional to global scale catastrophe and climatic perturbations \citep{Too1997}.\\

 As for Venus, Mars and the Moon, the largest crater generated
 by our simulated V-NEAs on each of this bodies is 17 km, 75 km and 6 km
 in diameter, respectively (Table 3).

\subsection{Typical orbits of the impactors}
The orbits of the impactors for the terrestrial planets analysed in this study, apart from Mars could appear like cometary orbits. 
In fact, the Tisserand parameter \citep{Opi1976} with respect to Jupiter can be less than 3 (Table ~\ref{craterele}).
\begin{equation}
T_P=\frac{a_p}{a}+2\sqrt{\frac{a}{a_p}(1-e^2)}\cos i,
\end{equation}

here $P$ stands for planet, $a_p$ is the semi-major axis of the planet, $a$, $e$ and $i$ are the osculating semimajor axis,
 eccentricity and inclination, respectively, of the asteroid for which we want to compute this parameter.

Curiously, some clones in our simulation, before they impact the Earth, possess the similar range of elements as the Bunburra Rockhole (BR) meteorite (associated to V-types due to its basaltic composition: the oxygen isotopic composition relative 
to eucrites and diogenites, meteorites associated with 4 Vesta, the angrite group, and anomalous eucrite-like basaltic achondirites)
 that fell in Western Australia on 21 July 2007 with an initial velocity of 13.4 km/s with $a=0.851$ au, $e=0.245$ and $i=9^\circ$ \citep{Bla2009}.
 In particular, 2 clones have similar elements at $\sim 1$ LD: 
   a clone of asteroid 1999 CV$_8$ with $a=0.804$ au, $e=0.415$ and $i=14^\circ$,
 and a clone of asteroid 2000 HA$_{24}$ with $a=0.882$ au, $e=0.307$ and $i=4^\circ$.
 The former has an entry velocity (at 100 km of altitude) of 16.6 km/s and the latter 15.0 km/s.

The range of orbital osculating elements for V-type NEAs before collision with terrestrial planets are shown in Table ~\ref{craterele}. 
Their Tisserand parameter shows that they could be mistaken for comets when $T_j<3$. This situation might not arise for other types of
 asteroids that originate from other regions of the solar system or when NEAs belong to a different cluster (e.g., different asteroid
 families with different orbital osculating elements, and thus Tisserand parameter). 
In addition, any region (an asteroid family, in particular) has its own probability of close encounters and impacts.
 For example, the E-types (from Hungarias) can have their semi-major axis at values much higher than that of V-types
 (see Table 7 in \citet{Gal2013} or different impact probabilities, see Table~\ref{impactcom}).

\begin{table*}[!h]
\begin{center}
\caption{V-NEA's impactors elements at the onset of atmospheric entry (100 km altitude).
The columns represent the planet, the osculating elements and Tisserand parameter.
The upper line in each row represents the  average value 
 for each element of the impactor  ($\bar{a}$,  $\bar{e}$ and  $\bar{i}$), 
while the lower line represents the entire range of values
 (from minimum to maximum for each element):  $\Delta a$,  $\Delta e$ and  $\Delta i$.}

\begin{tabular}{lcccc}
\hline 
\hline
Planet& $\bar{a}$ [au] &  $\bar{e}$ &$\bar{i}[^\circ]$   &$\bar{T_J}$ \\
   &  $\Delta a$  [au] &  $\Delta e$ & $\Delta i[^\circ]$ & $\Delta T_J$ \\ 
\hline
\hline
Venus & 1.15  & 0.51& $13.3$  & 5.5 \\
  &  $0.59<a<2.49$  &  $0.11<e<0.93$ &  $0.8 <i<28.9$ & $2.4<T_j<9.3$\\
\hline
Earth & 1.23    & 0.44  & $19.7$  & 5.0\\
 & $0.73<a<1.97$  &  $0.11<e<0.91$ &  $0.8 <i<50.2$ &  $3.0<T_j<7.6$\\
\hline
Moon & 1.30   & 0.46 & $8.4$  & 5.2 \\
 &  $0.78<a<2.05$  &  $0.38<e<0.60$ & $3.1 <i<16.5$ &  $3.0<T_j<7.1$\\
\hline
Mars & 1.46   & 0.60 & $21.5$  & 4.0\\
 &  $1.27<a<1.84$  &  $0.38<e<0.89$ & $6.3 <i<49.4$ &  $3.3<T_j<4.5$\\
\hline
\hline
\end{tabular}
	\label{craterele}		% notice the second label for counting
\end{center}
\end{table*}

 \subsection{Other end states}
The V-NEAs which do not collide with the planets in the NEA-region, have different fates; there are
  V-NEAs capable of escaping the inner solar system, hitting the Sun or even
 Jupiter (even if very seldom among V-Types: less than 0.1\% of all
 the integrated clones).
 Asteroids with $a\gas2.5$ au
 (J3:1 is at 2.5 au) preferentially migrate to the outer solar system or are ejected onto a hyperbolic orbit (e.g., asteroid 1999 RB$_{32}$).
 If the eccentricity is sufficiently large, the V-NEAs can become Jupiter-crossers and subsequently scattered outward by the giant planet.
 At that point, their dynamics becomes similar to that of Jupiter-Family comets
  \citep{Mic2005} with a $T_j<3$.

 Only those V-NEAs that are extracted 
 from the resonance by Mars or the Earth and rapidly transported on a low eccentricity path to smaller semi-major axes
 can escape the scattering action of Jupiter.
 However, such evolution is increasingly unlikely as the initial $a$ takes on
 larger and larger values.
Conversely, the bodies on orbits with $a\approx2.5$ au or less
 do not approach Jupiter even if $e\approx 1$.
Most of the V-NEAs with higher probability of colliding with the Sun
 or escaping are those initially in more chaotic regions
 ( i.e., between the Tisserand curves)
 and those in strong resonances (i.e., $\nu_6$, which influence V-NEAs such as 1997 $GL_3$,
 Verenia and 1980 WF (see Figure~\ref{Vlocation} and Table \ref{single}).
 
The total mortality or loss of the V-NEAs in 10 Myr is  43.3\%, of those
  5.3\% impacts with terrestrial planets, 26.4\%  with the Sun and 11.4\% escape (among these are also impacts with planets
 not considered in this study, i.e. a clone of 1997GL$_3 $ which has an impact with Jupiter), see Table \ref{single}.  
If we assume that the entire NEA population, including the V-NEAs as a subgroup, is in steady-state for at least the last 3 Gyr \citep{Bot2002,Mic2005}, then the lifetime required to drive the influx of new bodies to replenish the population is 23.1 Myr. 
In fact, the total number of clones impacting (461) and escaping (167) in 10 Myr is $\sim$43 \% of the grand-total.
 Therefore, $\sim$57\% of V-NEAs can survive in 10 Myr.

\noindent

\section{Conclusions}
\label{finallyMario}

In this work, we examined if present day V-NEAs can impact Venus, Earth, Moon and Mars) within the time span of the next 10 million years (Myr). 
The conclusions of our work are as follows:

(i) The known V-NEAs capable of causing damage on the terrestrial planets in the form of impacts are initially in high orbital inclinations,
 $i \leq 2$ and of kilometer size. Within 10 million years, about 91\% of V-NEAs are potentially PHAs for the Earth. If we extend the definition of PHAs 

 for the Earth to another planet (with a scaled distance for the Hill sphere for the relative planet), 
 there are approximately 75\% PHAs for Venus, 79\% for Mars and 58\% for the Moon. Compared to the Earth, the Moon
 has nearly half of close encounters. Some close encounters with the terrestrial planets,
 as opposed to Jupiter or the Sun, can expel the asteroids into the outer solar system, eventually sending them into highly eccentric or even retrogade orbits. The asteroids with the highest probability to collide with Venus and the 
 Moon are  1992 FE  and 2003 FU$_3$, respectively, and with the Earth are
 1996 EN and 2000 HA$_{24}$. The only V-NEA that does not produce any
 planetary impacts in the next 10 My is Magellan. 
 
(ii) The average encounter speed for Venus is higher than that for the other planets, as expected. 
 Additionally, Venus encounter speeds are on average greater than that for most of the intact asteroids, in particular Hungaria-NEAs,
 typically E-types. The average impact velocity  with the Earth deviate
 slightly from the corresponding value of  the average velocity of the entire  population of NEAs as terrestrial impactors.
 However, Mars experiences V-NEA impacts with a higher average velocity than typically expected by other types of asteroids. 
This could be either a statistical bias, due to the low number of impacts, or caused by orbits that are more eccentric than average.
In addition to close encounters, which are a key factor in changing the orbits, are also the resonances, especially for V-NEAs with major semi-axes
 larger than the orbit of Mars. Therefore, these initially undergo fewer close encounters.
Examples of V-NEAs perturbed by resonances are 1999 RB$_{32}$ (J3:1),
 1994 LX and Tjelvar (M4:3), Midas (J3:2), 1980 WF and 1997 GL$_3$ ($\nu_6$).
There are expections among typical V-type fates, e.g. some of them can survive for longer time than 10 Myr, i.e.,
it appears that the asteroid Magellan is less affected by resonances  and close encounters. 
Magellan clones never meet Venus and have very low chances to encounter
 the  Earth, the Moon and Mars. In fact, Magellan has a higher perihelion than the average V-NEAs and the highest MOID value for Venus.

(iii) The V-NEAs mainly impact the Earth: there is 3\% of such V-NEAs in 10 million years.
 Only about 1.5\% will collide with Venus and even less with Mars and the Moon, as opposed to E-types (Hungarias)
 which preferentially impact Venus, then Mars and ultimately the Earth. 
 
(iv) Impacts with the Earth, according to our sample, occur every $\sim$12 million years and have the potential to cause disastrous effects on regional to global scale, producing craters as large as 30 km in diameter and releasing kinetic energy of as much as 3 Mt. 
This energy is almost 6 million times
 greater than the energy released during the Chelyabinsk event in 2013. 
Venus, Mars and the Moon will experience impacts with V-NEAs every 22 Myr, 125 Myr and 168 Myr, respectively.

(v) Among the craters that are likely to be formed by basaltic NEAs, we find four possible crater candidates on the Earth, two for which the confirmation is still pending and two for which the impactor comosition was deemed to be of basaltic origin. If the impactors' analysis are to be confirmed for a basaltic impactor, the Ries crater (24 km diameter) in Germany, and the El'gygytgyn crater (18 km diameter) in Russia should be added to the crater candidates list. The two craters with confirmed basaltic impactor are the Strangways crater (24 km diameter) in Australia, the Nicholson crater (12.5 km diameter) in Canada.

(vi) For Mars, Venus and the Moon, for the craters with the largest diameters (e.g., 75 km, 17 km and 6 km, respectively) found in this study, 
it is not inconceivable that they could have been created by basaltic NEAs.

(vii) Comparing this work with \citet{Cha2004}, it appears
 V-NEAs have a higher probability of colliding with the Earth compared to other types of NEAs. 
 Then, the V-NEAs with a higher probability of being expelled or impacting the Sun are typically in those chaotic regions:
  orbits between Earth and Venus or in-between Tisserand curves, and regions influenced by secular resonances such as the $\nu_6$.
If we consider the NEAs in steady-state and therefore the V-NEAs (at least until the Vesta family has formed),
 the time needed to replenish them is equal
 to 21.3 million years (Vesta family is much older, at least 1 Gyr, see \citet{Car2016}.

(viii) The mortality or loss rate in 10 million years is equal to $\sim$43\%:
 5.3\% impacts with terrestrial planets, 26.4\% impacts with the Sun and 11.4\% escapers. A not significant number of V-NEAs can have impacts
 even with Jupiter ($<0.1$\% in 10 Myr).

(ix) Concerning the statistical study on the elements osculating before the impact (100 km above the Earth's surface), interestingly, we find that some V-NEAs have orbits similar to that of the BR meteorite. \\

Finally, we state that this work can be extended to discriminate asteroid families among NEAs (since they can have different osculating elements and interactions with the planets), especially if coupled with observations of their physical charatestics (e.g. spectral type).

\acknowledgements
MAG was supported by the FWF: Project J-3588-N27 ``NEOs, impacts, origins and stable orbits'' and, he thanks Prof. B. A. Ivanov and Prof. H. J. Melosh for the suggestions regarding impacts material, Prof. V. Carruba,  Prof. R. Dvorak, Dr. S. Eggl,  Dr. G. Osinski,  Prof. J. Souchay, Dr. D. Souami and Prof. P. Wiegert for further suggestions on this project.
EAS gratefully acknowledges the developers of iSALE-2D (www.isale-code.de), the simulation code used in our research, including G. Collins, K. Wunnemann, B. A. Ivanov, D. Elbeshausen and H. J. Melosh.

% Example of using BiBTeX (plus natbib):
% For details see \cite{1999MNRAS.309..731B},
% \cite{1893PASP....5..204C},
% \cite{2008IAUS..252...75L}. It has been demonstrated that this
% is important \citep{2012AN....333..663S}.

% Use this code if you wish to generate your bibliography with BibTeX;
% please replace first the string "an-demo" below with the name(s) of
% the BibTeX data base(s) you want to use.
% The resulting bibliography-output (the contents of the .bbl file)
% must be pasted into this file before submission.
% 
%\bibliographystyle{an}
%\bibliography{an-demo}
% 
% Replace the following example bibliography with your references
% before submission:

%\bibliographystyle{plainnat} % that's for the display
%\bibliography{vtypeOK.bib}

\end{document}